\documentclass[10pt,a4paper]{article}
\usepackage[textwidth=153mm,textheight=250mm,centering]{geometry}
\usepackage{graphicx}
\usepackage{xcolor}
\usepackage{booktabs}
\usepackage{array}
\usepackage{longtable}
\usepackage{float}
\usepackage{amsmath}
\usepackage{textcomp}
\usepackage[font=small,labelfont=bf,labelsep=period]{caption}
\usepackage[colorlinks=true,allcolors=blue]{hyperref}

\newcommand{\uV}{\,\textmu V}
\raggedright
\let\oldtabular\tabular
\renewcommand{\tabular}{\footnotesize\oldtabular}

\newcommand{\articletype}[1]{}
\renewcommand{\title}[1]{{\LARGE\bfseries\raggedright #1\par}\vspace{4mm}}
\renewcommand{\author}[1]{{\normalsize\raggedright #1\par}\vspace{2mm}}
\newcommand{\affil}[1]{{\footnotesize\raggedright #1\par}}
\newcommand{\email}[1]{\vspace{2mm}{\footnotesize\raggedright\textbf{E-mail:} #1\par}\vspace{2mm}}
\newcommand{\keywords}[1]{{\footnotesize\raggedright\textbf{Keywords:} #1\par}}
\newcommand{\orcid}[1]{\href{https://orcid.org/#1}{\includegraphics[width=8pt]{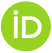}}}
\renewenvironment{abstract}{\vspace{4mm}{\color{gray}\hrule}\vspace{2mm}%
  \noindent{\bfseries Abstract}\par\vspace{1mm}\noindent\ignorespaces}%
  {\par\vspace{2mm}{\color{gray}\hrule}\vspace{4mm}}
\newcommand{\ack}[1]{\section*{Acknowledgments}#1}
\newcommand{\funding}[1]{\section*{Funding}#1}
\newcommand{\data}[1]{\section*{Data availability}#1}
\newcommand{\roles}[1]{\section*{Author contributions}#1}
\newcommand{\suppdata}[1]{\section*{Supplementary data}#1}

\begin{document}

\articletype{Paper}

\title{Cross-subject decoding of imagined speech from scalp EEG: a leakage-controlled evaluation on three public corpora}

\author{Blessing Isoyiza Adeika$^{1,*}$\orcid{0009-0007-5600-8051} and Onyema Osuagwu$^1$}

\affil{$^1$Morgan State University, Baltimore, Maryland, United States of America}

\affil{$^*$Author to whom any correspondence should be addressed.}

\email{blade3@morgan.edu}


\begin{abstract}
\textit{Objective.} A decoder of imagined speech that works for a new user without calibration would make a non-invasive speech interface practical. We tested whether imagined-speech content recorded with scalp electroencephalography (EEG) transfers across people, and examined why accuracy within a single person remains low.
\textit{Approach.} We analyzed three public corpora, ds003626, KaraOne and Track~3 of the 2020 International Brain--Computer Interface Competition, under a protocol that enforces split integrity by assertion and reports exact binomial thresholds, group tests and bootstrap confidence intervals. We trained decoders on each participant and tested them on every other, compared representational geometry across participants without a classifier, applied subject-level alignment, and tested five explanations for the within-subject ceiling.
\textit{Main results.} Across 482 ordered pairs of participants, no pair survived correction for multiple comparisons; mean pair-transfer accuracy lay within 0.006 of chance on every corpus, while every within-subject reference lay above it. Imagery against pre-stimulus rest generalized at about 0.61 after artifact and elapsed-time controls, but its decoder labelled the interval preceding imagery as imagery as often as imagery itself, so the contrast reflects a post-stimulus state, not imagery. Within subject, only the competition corpus decoded above threshold for every participant (0.3147; chance 0.200); syllable count explained little of this, but its frequency profile failed an artifact audit.
\textit{Significance.} In these corpora, imagined-speech content did not transfer between people through any representation we examined. Trial count per class was the only variable that ordered the corpora, which favors larger single-condition datasets over further model development.
\end{abstract}

\section{Introduction}

Most brain--computer interfaces (BCIs) must be calibrated for each user, because the signals they rely on differ between people and between sessions \cite{wolpaw2002,saha2020}. Transfer learning aims to reduce that burden by reusing data from other users \cite{jayaram2016,wu2022}. For speech the stakes are high. Intracranial speech neuroprostheses now reach high accuracy after short calibration \cite{card2024,silva2024}, whereas scalp electroencephalography (EEG), the only modality that could reach users without surgery, has produced imagined-word decoding that remains mostly at or near chance when corpora are re-evaluated under a common protocol \cite{panachakel2021,lopezbernal2022,vandyck2026}. A decoder that generalized across people would indicate that imagined speech leaves a shared, measurable trace at the scalp; one that did not would locate the difficulty in the signal itself, not in the absence of calibration data.

Direct tests of cross-subject transfer for imagined speech are scarce. A recent benchmark of pretrained EEG models on the 2020 BCI Competition data found every model at chance when the test participant was held out \cite{khanday2026}, and a cross-dataset evaluation found within-category decoding mostly at or near chance even within subject \cite{vandyck2026}. A failure of leave-one-subject-out (LOSO) decoding is, however, ambiguous. Participants may share a representational geometry that pooling across them blurs into noise, in which case selecting compatible source participants or aligning their data could succeed \cite{he2020,zanini2018,rodrigues2019}. Alternatively, nothing may be shared, in which case no alignment or adversarial adaptation \cite{ganin2016} can create transfer, and attempts to force it risk negative transfer \cite{zhang2023}. Distinguishing the two requires testing transfer between individual participants and comparing representations without relying on a classifier.

The within-subject ceiling raises a parallel question. Accuracies close to chance within one person admit several explanations: too many features for the available trials, a representation that misses cross-channel structure, an analysis window that dilutes a brief effect, too little training data, or a model that is too weak. Each makes a different prediction, and following the logic of strong inference \cite{platt1964}, we gave each an experiment that could overturn it.

We report these tests on three public corpora under a single leakage-controlled protocol. The protocol, its artifact audit of the positive controls available in these corpora, and the detection floors that bound the null results reported here are described in a companion paper \cite{companionA1}. Here we ask whether imagined-speech content transfers between participants, whether any shared structure is detectable without a classifier, whether subject-level alignment changes the answer, why within-subject accuracy is low, and whether phonological features generalize to prompts not seen in training.

\section{Related work}

\subsection{Reported performance on these corpora}

Each of the three corpora comes with a published benchmark. KaraOne was designed around phonological minimal pairs, and its authors reported up to 90\% accuracy for binary consonant-versus-vowel classification of imagined phonemes, within subject, using deep belief networks \cite{zhao2015}; a hierarchical deep network later reached a mean of 83.42\% over six binary phonological contrasts \cite{saha2019}. On the four-class inner-speech task of ds003626, EEGNet reached 29.7\% \cite{vandenberg2021} and a bidirectional recurrent network 36.1\% \cite{gasparini2022}, against a chance level of 25\%. In Track~3 of the 2020 BCI Competition, which comprises five imagined phrases, the leading entries reached 82.6\% and 75.5\% \cite{jeong2022}. Pooling pronounced and imagined trials has been reported to improve within-subject decoding of some imagined word pairs \cite{alonsovazquez2025}. Reviews of the field describe a wide range of results that are hard to compare across studies because paradigms, vocabularies and evaluation schemes differ \cite{panachakel2021,lopezbernal2022,jin2025,su2025}.

Re-evaluations that apply a common protocol report much lower figures. When seven decoders were compared on nine corpora under one nested cross-validation scheme, decoding within a category of imagined speech was mostly indistinguishable from chance, and the best pairwise balanced accuracy, 62.7\%, was obtained on the 2020 competition data \cite{vandyck2026}. On the same data, a benchmark of five models, including two pretrained EEG foundation models, reported within-subject balanced accuracies of 25.7 to 30.2\% for five classes and chance for every model under LOSO evaluation \cite{khanday2026}.

\subsection{Why reported accuracies diverge}

Several mechanisms can raise accuracy without speech decoding. When segments of one recording are distributed across training and test sets, a classifier can learn to recognize the recording instead of the class, and segment-wise splitting has been shown to inflate deep-learning accuracy on EEG substantially \cite{brookshire2024}. Feature selection or model tuning performed outside cross-validation, and selection on the test data themselves, bias results in the same direction \cite{kriegeskorte2009,whelan2014,kaufman2012,kapoor2023}. In block designs, slow drift in the recording becomes a proxy for the experimental condition \cite{li2021}, a hazard that applies to the condition axis of ds003626 \cite{companionA1}. Confounds correlated with the label are a further route: in KaraOne, the prompt type alone, word versus phoneme, decodes at 0.8371 with all 14 participants above threshold in our data, so any contrast whose classes differ in their proportion of words can succeed on duration and prompt type without phonological information. Restricting KaraOne to its seven phonemes is the corresponding control: it reduced the effect in the 2\,s interval that precedes imagery by 53\% and removed an imagined-speech effect entirely (a gain over chance of $+$0.0288, $p$ = 0.0011, became $-$0.0015 with no participant above threshold, $p$ = 0.514). Finally, with the small test sets typical of these corpora, theoretical chance levels are misleading and single cross-validated estimates carry wide error bars \cite{combrisson2015,varoquaux2018}.

\subsection{Cross-subject transfer}

Transfer learning in BCIs has been reviewed extensively \cite{jayaram2016,wu2022}, and inter-subject variability is recognized as its central obstacle \cite{saha2020}. Euclidean alignment whitens each participant's trials by their mean spatial covariance \cite{he2020,wu2025}, Riemannian alignment re-centers each participant's covariance matrices at their geometric mean \cite{zanini2018}, and Procrustes analysis extends re-centering with stretching and rotation \cite{rodrigues2019}. Domain-adversarial training learns features that predict the task but not the domain \cite{ganin2016}. These methods presuppose that participants share class structure that differs only by a transformation; when that premise fails, transfer can reduce accuracy instead of improving it \cite{zhang2023}. Representational similarity analysis tests the premise directly by comparing the geometry of class representations across individuals without a decoder \cite{kriegeskorte2008}.

\subsection{Representations beyond band power}

Covariance-based classifiers on the Riemannian manifold are among the most reliable decoders for small EEG datasets \cite{barachant2012,congedo2017,lotte2018}. Convolutional networks designed for EEG, such as EEGNet and ShallowConvNet, learn spatial and temporal filters from raw signals \cite{lawhern2018,schirrmeister2017}, and attention-based architectures such as EEG Conformer extend them \cite{song2023}; systematic reviews note, however, that reported gains are often hard to reproduce \cite{roy2019}. Foundation models pretrained on large EEG collections aim to transfer across tasks and people \cite{kostas2021,jiang2024}, but did not show a consistent advantage on speech \cite{khanday2026}. Intracranial recordings point to low-frequency power and cross-frequency coupling as informative features for imagined speech \cite{proix2022}, which motivates phase-based measures such as the weighted phase-lag index and phase--amplitude coupling \cite{vinck2011,tort2010,canolty2006}, provided they are computed in forms that are robust to volume conduction and the reference \cite{bastos2016,nolte2004}.

\section{Materials and methods}

\subsection{Corpora and tasks}

Table~\ref{tab:tasks} lists the decoding tasks. The recordings, their trial structure, the preprocessing and the corrections required to read ds003626 correctly are described in the companion paper \cite{companionA1}; in brief, ds003626 is analyzed with the corrected condition mapping (code~1 is inner speech) over the 1.0--3.5\,s action interval \cite{nieto2022,nieto2022data}, KaraOne over the imagery interval using the 62 scalp channels re-referenced to their own mean \cite{zhao2015}, and the competition data over the whole 3.1\,s epoch \cite{jeong2022}. Each KaraOne trial passes through four states: rest, a stimulus state in which the prompt is shown and played and which ends with a 2\,s period in which participants move their articulators into position, imagined speech, and speaking aloud \cite{zhao2015}. For comparison we also analyzed the 2.0\,s interval between rest and imagery, which we call the pre-imagery interval; it contains perception of the prompt and preparation to articulate it, but neither imagery nor speech. The recordings' segment annotations list this interval among their speaking segments, so it is easily mistaken for spoken speech \cite{companionA1}. KaraOne's spoken productions, which follow imagery, were not analyzed. Unless stated otherwise, data were band-pass filtered at 1--45\,Hz and resampled to 128\,Hz.

\begin{table}[t]
\caption{Decoding tasks. Trials per class are per participant. For ds003626 the inner-speech condition is the primary task and the pronounced condition is reported for comparison.}
\label{tab:tasks}
\centering
\begin{tabular}{@{}llccl@{}}
\toprule
Corpus & Task & Classes & Trials per class & Participants \\
\midrule
2020 BCI Competition, Track 3 & imagined phrases & 5 & 80 & 15 \\
ds003626 & inner-speech words & 4 & about 56 & 10 \\
ds003626 & pronounced words & 4 & about 28 & 10 \\
KaraOne & imagined words & 4 & about 12 & 14 \\
KaraOne & phonemes, imagined and pre-imagery & 7 & varies & 14 \\
KaraOne & binary phonological contrasts & 2 & varies & 14 \\
KaraOne & imagery versus rest & 2 & varies & 14 \\
\bottomrule
\end{tabular}
\end{table}

\subsection{Protocols and decoders}

Every split was generated by one module and checked by an assertion that halts the run if a trial appears in more than one set, if the test participant appears in training data under LOSO, or if a test class is absent from training \cite{companionA1}. Within subject we report stratified 5-fold cross-validation, which pools sessions and is reported because the literature uses it, together with a held-out protocol that tests on an unseen session (ds003626) or an unseen contiguous block of trials (KaraOne and the competition corpus). Across participants we report LOSO evaluation. No selection signal was taken from test data.

The reference decoder is L2-regularized multinomial logistic regression on the logarithm of band power per channel in six bands from 1 to 45\,Hz, with the regularization fixed \cite{pedregosa2011}. We chose this feature set before running any analysis and use it as the fixed comparator against which every alternative is judged. Comparators were EEGNet with 16 temporal filters and a depth multiplier of 4 \cite{lawhern2018}; Riemannian classifiers (minimum distance to mean, its geodesically filtered variant, and tangent-space logistic regression) with optional principal component reduction and per-participant re-centering \cite{barachant2010,barachant2012,congedo2017,pyriemann}; ShallowFBCSPNet and EEG Conformer from the Braindecode library, version 1.2.0 \cite{braindecode,schirrmeister2017,song2023}; and logistic regression on phase, connectivity and coupling features \cite{vinck2011,tort2010}. The two Braindecode models were used at their published default hyperparameters, except that ShallowFBCSPNet's dropout was set to 0.3, the value used for every network in this study; on the ds003626 input shape they have 208,324 and 488,132 parameters. All networks were trained with the Adam optimizer, a learning rate of $10^{-3}$, weight decay of $10^{-4}$, batches of 32 and at most 100 epochs, with early stopping on a validation set drawn from training data. The detection floors of these decoders, measured by injecting effects of known size into the same recordings, are reported in the companion paper \cite{companionA1}.

\subsection{Analyses}

\textit{Content decoding.} Each corpus was decoded with the reference decoder under all three protocols. Pairwise word separability was measured on matched 128\,Hz, 1--45\,Hz data in three bands (8--13, 8--30 and 30--45\,Hz); for the competition corpus each pair was also decoded in a 62--100\,Hz band and in the 8--30 and 62--100\,Hz bands combined, from a 256\,Hz version of the data. To test whether utterance length could explain separability on the competition corpus, its pairs were grouped by whether the two phrases have the same number of syllables (one for ``stop'' and ``yes'', two for ``hello'', ``help me'' and ``thank you''), which gives four matched and six mismatched pairs; matched and mismatched accuracies were averaged per participant and compared with a paired one-sided Wilcoxon test.

\textit{Subject-level alignment.} Euclidean alignment was computed transductively from all of a participant's trials and inductively from their first 10, 20 or 40 trials, and Riemannian alignment re-centered each participant's covariances at their Riemannian mean. Every alignment was fitted per participant from that participant's own unlabeled trials, including the held-out participant, with no shared reference; these runs are flagged as using unlabeled target data.

\textit{Pair transfer and representational similarity.} For every ordered pair of participants we trained on one and tested on the other, so that the test participant contributed nothing to fitting. The within-subject reference on the diagonal was computed by 5-fold cross-validation. For the classifier-free comparison, each participant's representational dissimilarity matrix was built from Riemannian distances between class-mean covariance matrices \cite{kriegeskorte2008,congedo2017}, and matrices were compared across participant pairs with Spearman correlation. The null distribution permuted class labels within each participant 200 times, which destroys class structure while preserving each participant's covariance scale and channel geometry.

\textit{State versus content.} On KaraOne we decoded imagery against rest, using the re-referenced scalp channels. Because rest always precedes the stimulus and imagery follows it, the contrast is entangled with elapsed time; we therefore compared rest and imagery windows at different separations and trained and tested on different halves of each session. Splits were grouped by trial so that a trial's rest and imagery segments never fell into different folds. To separate imagery from what follows the stimulus, we cut the pre-imagery interval from the same recordings with the same preprocessing, taking it directly as the gap between the end of rest and the start of imagery. We decoded each pair of the three windows (rest, $-$2.0 to 0.0\,s; pre-imagery, 0.0 to $+$2.0\,s; imagery, from a median of $+$2.03\,s for 2.0\,s) under LOSO, and applied the rest-versus-imagery decoder, trained on the other participants, to all three windows of the held-out participant. A decoder that reads imagery should label the pre-imagery interval as rest.

\textit{Explanations for the within-subject ceiling.} We tested five explanations against the reference decoder, each with a paired per-participant test and Benjamini--Hochberg correction within its family. (i) Upstream failure: the same decoder was applied to the three-way condition contrast of ds003626, balanced per participant and session. (ii) Dimensionality: a single 8--30\,Hz band, principal component projections to 16, 32 and 64 components fitted on training data, an anatomical speech-motor region, and cropped training that multiplies training rows fivefold. (iii) Missing cross-channel structure: 23 Riemannian configurations combining spatial reduction, filter banks, re-centering and the three classifiers. (iv) The analysis window: 500\,ms windows in 50\,ms steps, tested with a cluster-based permutation over time \cite{maris2007}. (v) Training data: adding the pronounced and visualized trials of ds003626, restricted to sessions already in the training fold, to the inner-speech training set. We also ran the two Braindecode architectures under both protocols.

\textit{Held-out prompts.} For the five KaraOne phonological contrasts (nasal, bilabial, /iy/, /uw/, consonant versus vowel), each test fold held out one prompt from each side of the contrast, giving 112 valid hold-out combinations. Each contrast was run over all 11 prompts and, where a positive prompt remained for training, over the seven phonemes alone, both with the test participant seen in training and with the test participant held out.

\textit{Phase features.} Connectivity (signed weighted phase-lag index), theta--beta phase--amplitude coupling and phase at a fixed latency were computed per trial and band for a language region set and a central sensorimotor region on all three corpora and both protocols. The first two families were validated on injected ground truth before use \cite{companionA1}; the third was not and is treated as exploratory.

\subsection{Statistical analysis}

Each result carries four statistics: the smallest accuracy that is significant for a single participant under the exact binomial distribution at $\alpha = 0.05$ \cite{combrisson2015} and the number of participants who reach it; a one-sided Wilcoxon signed-rank test of participant accuracies against chance \cite{wilcoxon1945}; and a 95\% confidence interval (CI) on the mean from a bias-corrected and accelerated bootstrap \cite{efron1987}. Where several related tests were made, the false discovery rate (FDR) was controlled with the Benjamini--Hochberg procedure \cite{benjamini1995}; $q$ denotes the adjusted $p$-value. Every positive result reported here was also checked with a shuffle-label control in which labels were permuted within each participant and the identical pipeline re-run (table~\ref{tab:shuffle}); the result stands only if the null mean is not above chance and the observed value exceeds the 95th percentile of the null.

\begin{table}[t]
\caption{Shuffle-label controls for the positive results reported in this paper. Accuracy is the value measured in this study. Labels were permuted within participant and the pipeline re-run 200 times, or 50 times for the network, whose every draw is a full retraining. No null draw reached the observed value in any row, so each empirical $p$ equals its minimum of $1/(n+1)$: 0.005 for 200 draws and 0.020 for 50.}
\label{tab:shuffle}
\centering
\begin{tabular}{@{}lcccc@{}}
\toprule
Result & Accuracy & Chance & Null mean & Null 95th pct \\
\midrule
KaraOne, imagery vs rest, 5.0\,s gap & 0.6475 & 0.500 & 0.4988 & 0.5133 \\
KaraOne, imagery vs rest, 2.0\,s gap & 0.6099 & 0.500 & 0.4998 & 0.5128 \\
KaraOne, rest vs pre-imagery & 0.6646 & 0.500 & 0.4996 & 0.5163 \\
KaraOne, pre-imagery vs imagery & 0.6122 & 0.500 & 0.4999 & 0.5142 \\
ds003626, three-way condition & 0.6065 & 0.333 & 0.3317 & 0.3453 \\
Competition, within subject, band power & 0.3147 & 0.200 & 0.1969 & 0.2057 \\
Competition, within subject, ShallowFBCSPNet & 0.2438 & 0.200 & 0.1959 & 0.2035 \\
\bottomrule
\end{tabular}
\end{table}

Two features of the null distributions matter for interpretation. Cross-subject nulls fall on chance (0.4988 to 0.4999 here, and 0.4999 on average across 40 further cells), whereas within-subject nulls fall slightly below it (0.3317 against 0.3333, and 0.1969 and 0.1959 against 0.2000, with a mean of 0.4910 against 0.500 across 40 further binary cells). Training on permuted labels within a participant does not simply remove signal; it teaches the model a mapping that is anti-correlated with the held-out fold. Empirical $p$-values are therefore anti-conservative within subject, and we use the binomial threshold and the group test to decide whether accuracy exceeds chance, reserving the shuffle control for the separate question of whether the pipeline manufactures signal. The three-way condition result is 0.6065 in table~\ref{tab:shuffle} and 0.6079 in section~\ref{sec:ceiling} because the control averages over folds, which weights sessions equally, whereas section~\ref{sec:ceiling} averages over participants.

\section{Results}

\subsection{Content decoding within and across participants}\label{sec:content}

Table~\ref{tab:content} reports the reference decoder for every corpus, condition and protocol. No cross-subject cell reached its per-participant threshold on any corpus; the largest number of participants above threshold under LOSO was two of 15, on the competition corpus. The imagined-phoneme cell on KaraOne gave a nominal group $p$ of 0.049 under LOSO with no participant above threshold, which does not survive correction across the table. Under LOSO, macro-averaged F1 fell below accuracy (0.1796 against 0.2063 on the competition corpus, and 0.2116 against 0.2590 on ds003626 inner speech), which indicates that the classifier distributed its predictions unevenly across classes, as expected when no class structure has been learned and predictions follow asymmetries in the training participants.

\begin{table}[t]
\caption{Content decoding with the reference decoder. All values were measured in this study; accuracy is the mean over participants. ``5-fold'' pools sessions and is optimistic; ``held out'' tests on an unseen session (ds003626) or an unseen block of trials (other corpora). Chance is 0.200 for the five phrases, 0.250 for the four-word tasks and 0.143 for the seven phonemes. Threshold is the exact binomial per-participant threshold at $\alpha = 0.05$; $n_{\geq}$ is the number of participants at or above it; $p$ is the one-sided Wilcoxon test against chance; CI is the bootstrap 95\% confidence interval.}
\label{tab:content}
\centering
\setlength{\tabcolsep}{3.5pt}
\begin{tabular}{@{}p{27mm}lllcccl@{}}
\toprule
Corpus and task & Protocol & Decoder & Accuracy & Macro-F1 & Threshold & $n_{\geq}$ & $p$ \;\; [95\% CI] \\
\midrule
Competition, 5 phrases & 5-fold & band power & 0.3273 & 0.3270 & 0.235 & 14/15 & 0.0003 [0.305, 0.349] \\
 & held out & band power & 0.3147 & 0.3142 & 0.235 & 15/15 & $<$0.0001 [0.294, 0.336] \\
 & LOSO & band power & 0.2063 & 0.1796 & 0.235 & 2/15 & 0.2567 [0.191, 0.222] \\[2pt]
ds003626, 4 words, inner & 5-fold & band power & 0.2595 & 0.2595 & 0.305 & 2/10 & 0.2695 [0.237, 0.288] \\
 & held out & band power & 0.2788 & 0.2729 & 0.305 & 2/10 & 0.0137 [0.261, 0.300] \\
 & LOSO & band power & 0.2590 & 0.2116 & 0.305 & 0/10 & 0.0947 [0.247, 0.270] \\[2pt]
ds003626, 4 words, pronounced$^{a}$ & 5-fold & EEGNet & 0.4231 & 0.4178 & 0.330 & 8/10 & 0.0020 [0.355, 0.492] \\
 & LOSO & EEGNet$^{b}$ & 0.2511 & 0.2466 & 0.330 & 1/10 & 0.6436 [0.223, 0.295] \\[2pt]
KaraOne, 4 words, imagined & 5-fold & band power & 0.2985 & 0.2992 & 0.367 & 2/14 & 0.0257 [0.247, 0.342] \\
 & held out & band power & 0.2583 & 0.2490 & 0.367 & 1/14 & 0.3468 [0.222, 0.291] \\
 & LOSO & band power & 0.2643 & 0.2088 & 0.367 & 1/14 & 0.1811 [0.240, 0.296] \\[2pt]
KaraOne, 7 phonemes, imagined$^{a}$ & 5-fold & phase & 0.1603 & 0.1587 & 0.221 & 3/14 & 0.1446 [0.140, 0.188] \\
 & LOSO & phase & 0.1570 & 0.1524 & 0.221 & 0/14 & 0.0490 [0.138, 0.173] \\[2pt]
KaraOne, 7 phonemes, pre-imagery$^{a}$ & 5-fold & EEGNet & 0.1666 & 0.1594 & 0.214 & 4/14 & 0.0502 [0.140, 0.192] \\
 & LOSO & EEGNet & 0.1661 & 0.1518 & 0.214 & 3/14 & 0.0885 [0.145, 0.194] \\
\bottomrule
\end{tabular}

\vspace{2pt}
\raggedright\footnotesize $^{a}$Taken from other analyses and shown for comparison only; these rows enter no test or conclusion. The pronounced 5-fold row and the pre-imagery rows are fixed configurations of the companion paper's artifact audit \cite{companionA1}; the pronounced LOSO row and the imagined-phoneme rows are the highest of the runs logged for that task, protocol and decoder. The pronounced condition and the KaraOne pre-imagery interval are dominated by muscle activity \cite{companionA1}. $^{b}$EEGNet with 8 temporal filters and a depth multiplier of 2, the only configuration run for this cell; every other EEGNet row uses 16 and 4. Balanced accuracy, which tracked accuracy in every row, and the run identifiers are given in supplementary tables~S1a and S4.
\end{table}

Inner speech in ds003626 did not decode reliably within subject either. With sessions pooled, accuracy was 0.2595 against a threshold of 0.305, with two of ten participants above it; with a session held out it was 0.2788, again with two of ten. The published results on this task, 29.7\% and 36.1\% \cite{vandenberg2021,gasparini2022}, were obtained under different protocols; the first is close to the single-participant threshold of 0.305 that applies to our test sets.

\subsection{Within-subject decoding on the competition corpus and pairwise separability}\label{sec:bcic}

\begin{figure}[t]
\centering
\includegraphics[width=\textwidth]{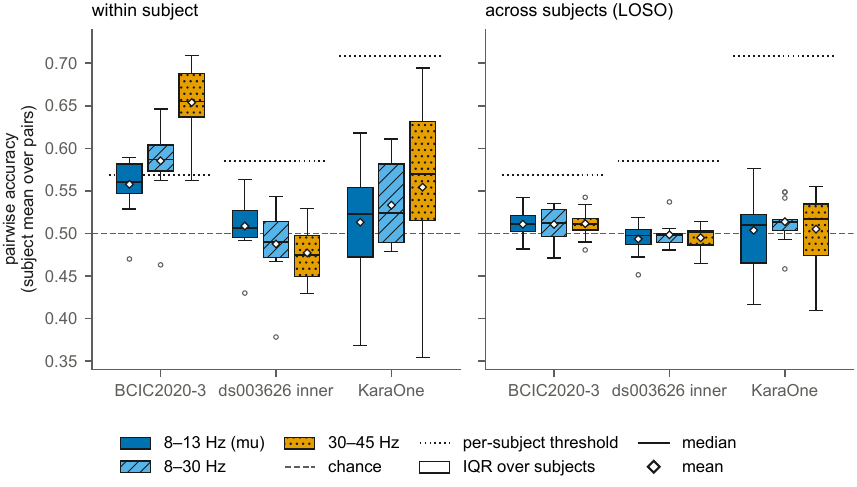}
\caption{Pairwise word separability by corpus, band and protocol. Each box summarizes one value per participant, that participant's accuracy averaged over the corpus's word pairs. The box spans the interquartile range, the center line is the median, the diamond is the mean, whiskers extend to 1.5 times the interquartile range and circles mark participants beyond them; fill and hatching identify the band. The dashed line is chance and the dotted segments each corpus's per-participant threshold.}
\label{fig:pairwise}
\end{figure}

\begin{figure}[t]
\centering
\includegraphics[width=\textwidth]{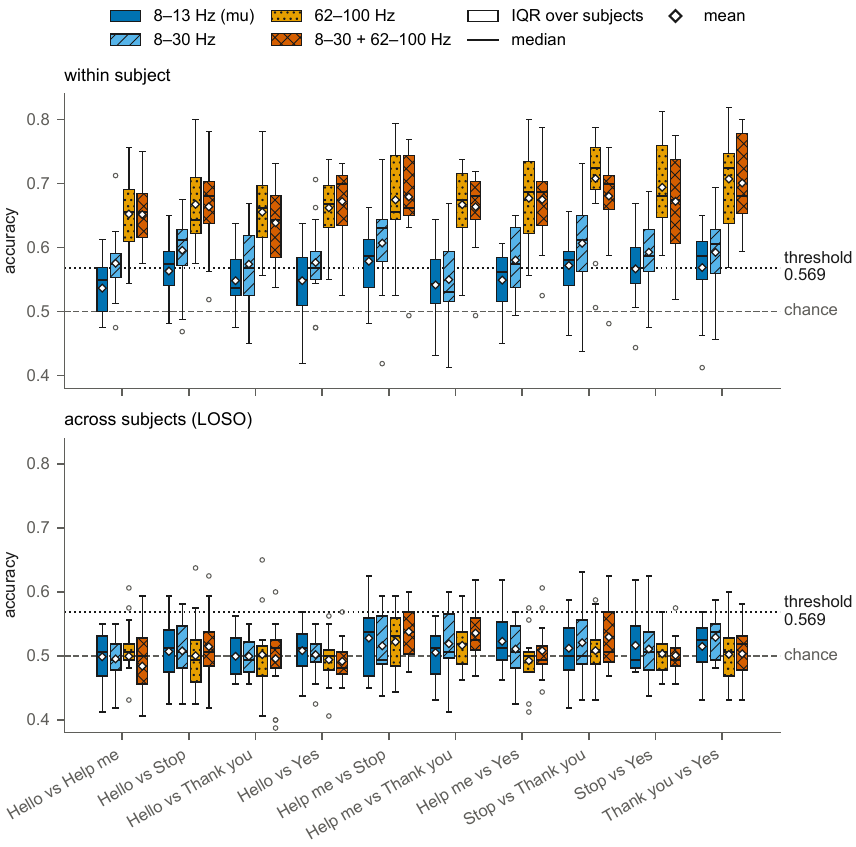}
\caption{Every phrase pair of the competition corpus, within subject (top) and across participants (bottom), in four bands. Each box summarizes the 15 participants, drawn as in figure~\ref{fig:pairwise}. The dashed line is chance and the dotted line the per-participant threshold (0.569).}
\label{fig:bcicpairs}
\end{figure}

The competition corpus is the exception within subject: 0.3273 with sessions pooled and 0.3147 with a block held out, with all 15 participants above threshold in the latter. Decomposed into its ten phrase pairs, every pair separated above chance within subject and survived FDR correction (table~\ref{tab:pairwise}, figures~\ref{fig:pairwise} and \ref{fig:bcicpairs}).

\begin{table}[t]
\caption{Pairwise word separability on matched 128\,Hz, 1--45\,Hz data. Chance is 0.500. Mean accuracy is over pairs; the CI treats participants as the unit, averaging each participant over pairs first. FDR correction is within each corpus.}
\label{tab:pairwise}
\centering
\begin{tabular}{@{}llccccc@{}}
\toprule
Corpus & Protocol & Band & Pairs & Accuracy & Threshold & Pairs surviving FDR \;\; [95\% CI] \\
\midrule
Competition & within & 8--13\,Hz & 10 & 0.5577 & 0.569 & 10/10 [0.538, 0.569] \\
 & within & 8--30\,Hz & 10 & 0.5853 & 0.569 & 10/10 [0.557, 0.600] \\
 & within & 30--45\,Hz & 10 & 0.6539 & 0.569 & 10/10 [0.631, 0.671] \\
 & LOSO & 8--13\,Hz & 10 & 0.5108 & 0.569 & 0/10 [0.503, 0.520] \\
 & LOSO & 8--30\,Hz & 10 & 0.5107 & 0.569 & 1/10 [0.500, 0.519] \\
 & LOSO & 30--45\,Hz & 10 & 0.5115 & 0.569 & 0/10 [0.504, 0.519] \\[2pt]
ds003626 inner & within & 8--13\,Hz & 6 & 0.5087 & 0.586 & 0/6 [0.483, 0.527] \\
 & within & 8--30\,Hz & 6 & 0.4877 & 0.586 & 0/6 [0.451, 0.509] \\
 & within & 30--45\,Hz & 6 & 0.4772 & 0.586 & 0/6 [0.459, 0.498] \\
 & LOSO & 8--13\,Hz & 6 & 0.4935 & 0.586 & 0/6 [0.479, 0.503] \\
 & LOSO & 8--30\,Hz & 6 & 0.4984 & 0.586 & 0/6 [0.492, 0.512] \\
 & LOSO & 30--45\,Hz & 6 & 0.4950 & 0.586 & 0/6 [0.484, 0.503] \\[2pt]
KaraOne & within & 8--13\,Hz & 6 & 0.5132 & 0.708 & 0/6 [0.478, 0.540] \\
 & within & 8--30\,Hz & 6 & 0.5332 & 0.708 & 0/6 [0.510, 0.560] \\
 & within & 30--45\,Hz & 6 & 0.5544 & 0.708 & 0/6 [0.499, 0.597] \\
 & LOSO & 8--13\,Hz & 6 & 0.5038 & 0.708 & 0/6 [0.480, 0.525] \\
 & LOSO & 8--30\,Hz & 6 & 0.5138 & 0.708 & 0/6 [0.501, 0.525] \\
 & LOSO & 30--45\,Hz & 6 & 0.5052 & 0.708 & 0/6 [0.481, 0.523] \\
\bottomrule
\end{tabular}
\end{table}

Three observations limit how this result can be read. First, the effect is present in the mu band (0.5577), which lies below the range where muscle activity dominates scalp recordings \cite{whitham2007,goncharova2003}, but it grows with frequency and is largest in the bands most exposed to muscle; the highest single cell on this corpus, 0.3815 for five-class decoding in 62--100\,Hz, belongs to the artifact audit of the companion paper \cite{companionA1}. Second, the effect does not replicate on the other corpora: the upper limits of their participant-level intervals in the mu band, 0.5268 for ds003626 and 0.5405 for KaraOne, lie below the competition corpus's point estimate of 0.5577. The intervals themselves overlap, so this does not exclude an effect of the same size, and differences in trial count, language and vocabulary would confound any direct comparison. Third, it does not transfer: in the 256\,Hz version of the data, under LOSO the effect collapsed in every band, from 0.5574 to 0.5114 in the mu band and from 0.6765 to 0.5042 in 62--100\,Hz, and only three of 40 cross-subject pair cells reached adjusted significance, each with a gain of about 0.03 and a minority of participants above threshold.

Because the within-subject accuracy exceeded what we expected for this problem, it was put through the verification checks described in the companion paper \cite{companionA1}, and it failed the physiological-plausibility check: its frequency profile concentrated at 30--45\,Hz, the range most exposed to muscle activity. The five phrases also differ in syllable count, so the duration confound that explains the KaraOne prompt-type effect could operate here. Length did contribute: within subject, pairs whose phrases differ in syllable count separated better than pairs that match, by 0.0151 to 0.0220 depending on the band ($q$ = 0.025 in three of four bands and 0.084 at 62--100\,Hz). It accounted for a small part of the effect, however. Matched pairs alone separated above chance in every band, from 0.5483 at 8--13\,Hz to 0.6673 at 62--100\,Hz ($p \leq 0.0007$), and their accuracy rose with frequency as that of all pairs did (supplementary table~S6). Under LOSO, matched pairs did not differ from chance ($p \geq 0.16$) and the two groups did not differ from each other ($p \geq 0.11$). Syllable count therefore does not explain the within-subject result, and the frequency profile, which persists in the matched pairs, remains the reason for caution. We therefore do not interpret the competition corpus's within-subject accuracy as evidence of imagined-speech content. It remains useful as an instrument: it is the only corpus in which every participant decodes above threshold, and therefore the one with the statistical power to show an improvement from a better representation if one existed (section~\ref{sec:ceiling}).

\subsection{Subject-level alignment}\label{sec:alignment}

Table~\ref{tab:alignment} reports every alignment cell under LOSO. None of the 18 cells reached its per-participant threshold, and none survived Benjamini--Hochberg correction across the family (smallest adjusted $p$ = 0.311). The largest change was $+$0.0390 on KaraOne, but the unaligned KaraOne baseline was 0.2390, 0.011 below chance, so the aligned cell lies 0.028 above chance; much of the apparent gain is a return toward chance from a below-chance start.

\begin{table}[t]
\caption{Cross-subject decoding with subject-level alignment (LOSO, band power, 8--30\,Hz). Every alignment is fitted per participant from that participant's unlabeled trials; ``inductive $K$'' uses only the first $K$ trials. Per-participant thresholds are 0.235 (competition), 0.305 (ds003626) and 0.367 (KaraOne). Change is relative to no alignment.}
\label{tab:alignment}
\centering
\begin{tabular}{@{}llcccc@{}}
\toprule
Corpus & Alignment & Accuracy & Change & $n_{\geq}$ & Wilcoxon $p$ \\
\midrule
Competition & none & 0.2080 & -- & 3/15 & 0.0956 \\
 & Euclidean, inductive 10 & 0.2063 & $-$0.0017 & 2/15 & 0.2131 \\
 & Euclidean, inductive 20 & 0.2122 & $+$0.0042 & 3/15 & 0.0625 \\
 & Euclidean, inductive 40 & 0.2055 & $-$0.0025 & 2/15 & 0.2534 \\
 & Euclidean, transductive & 0.2013 & $-$0.0067 & 1/15 & 0.4129 \\
 & Riemannian & 0.2052 & $-$0.0028 & 2/15 & 0.3661 \\[2pt]
ds003626 inner & none & 0.2453 & -- & 0/10 & 0.7051 \\
 & Euclidean, inductive 10 & 0.2402 & $-$0.0052 & 0/10 & 0.8516 \\
 & Euclidean, inductive 20 & 0.2440 & $-$0.0014 & 0/10 & 0.7422 \\
 & Euclidean, inductive 40 & 0.2490 & $+$0.0037 & 0/10 & 0.4531 \\
 & Euclidean, transductive & 0.2539 & $+$0.0086 & 0/10 & 0.2373 \\
 & Riemannian & 0.2462 & $+$0.0008 & 0/10 & 0.8242 \\[2pt]
KaraOne & none & 0.2390 & -- & 0/14 & 0.7357 \\
 & Euclidean, inductive 10 & 0.2771 & $+$0.0381 & 1/14 & 0.0691 \\
 & Euclidean, inductive 20 & 0.2631 & $+$0.0241 & 1/14 & 0.2662 \\
 & Euclidean, inductive 40 & 0.2753 & $+$0.0363 & 1/14 & 0.0494 \\
 & Euclidean, transductive & 0.2780 & $+$0.0390 & 2/14 & 0.0447 \\
 & Riemannian & 0.2771 & $+$0.0381 & 2/14 & 0.1044 \\
\bottomrule
\end{tabular}
\end{table}

\subsection{Pair transfer and representational similarity}\label{sec:shared}

\begin{figure}[t]
\centering
\includegraphics[width=\textwidth]{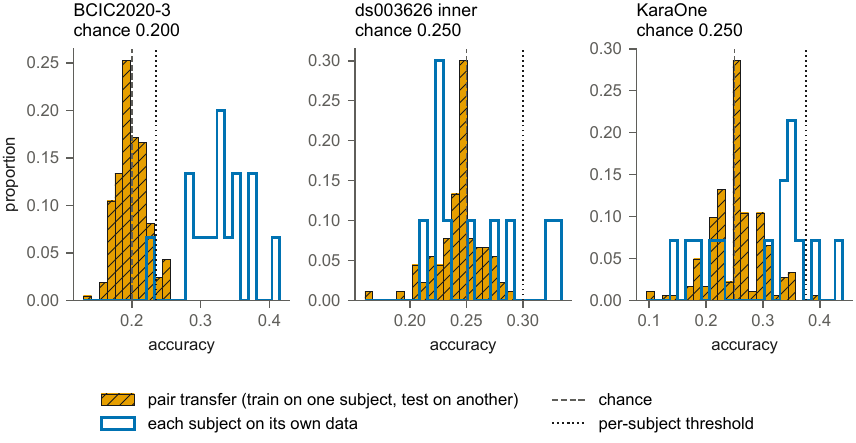}
\caption{Accuracy of decoders trained on one participant and tested on another (hatched bars, one value per ordered pair), against each participant's within-subject accuracy (outlined bars). Both are plotted as proportions so that pairs and participants share an axis. The dashed line is chance and the dotted line the per-participant threshold.}
\label{fig:transfer}
\end{figure}

Training on one participant and testing on another, for every ordered pair, gave 210 pairs on the competition corpus, 90 on ds003626 and 182 on KaraOne (table~\ref{tab:transfer}). Not one of the 482 pairs survived Benjamini--Hochberg correction within its corpus. Mean pair-transfer accuracy lay within 0.006 of chance on every corpus ($-$0.0004, $-$0.0052 and $+$0.0027), whereas every within-subject reference lay above chance ($+$0.1273, $+$0.0095 and $+$0.0486). On the competition corpus, where every participant decodes above threshold within subject, the distribution of pair-transfer accuracies is centered on chance and does not overlap most of the within-subject values (figure~\ref{fig:transfer}).

\begin{table}[t]
\caption{Transfer between individual participants and classifier-free representational similarity. The within-subject reference is 5-fold cross-validation computed in the pair-transfer run; for KaraOne it differs from table~\ref{tab:content} in the fourth decimal (0.2986 against 0.2985). Representational similarity is the mean Spearman correlation between participants' dissimilarity matrices over 105, 45 and 91 participant pairs, built from Riemannian distances between class-mean covariances, against a null that permutes labels within participant (200 permutations).}
\label{tab:transfer}
\centering
\setlength{\tabcolsep}{4pt}
\begin{tabular}{@{}lccccccccc@{}}
\toprule
 & & \multicolumn{5}{c}{Pair transfer} & \multicolumn{3}{c}{Representational similarity} \\
\cmidrule(lr){3-7}\cmidrule(lr){8-10}
Corpus & Pairs & Within & Transfer & Chance & $\geq$ threshold & FDR & Mean $r$ & Null 95th pct & $p$ \\
\midrule
Competition & 210 & 0.3273 & 0.1996 & 0.200 & 14/210 & 0/210 & $+$0.0403 & $+$0.0601 & 0.1095 \\
ds003626 inner & 90 & 0.2595 & 0.2448 & 0.250 & 0/90 & 0/90 & $-$0.0311 & $+$0.1176 & 0.5920 \\
KaraOne & 182 & 0.2986 & 0.2527 & 0.250 & 1/182 & 0/182 & $+$0.0179 & $+$0.1028 & 0.2985 \\
\bottomrule
\end{tabular}
\end{table}

Representational similarity between participants did not exceed its null on any corpus (table~\ref{tab:transfer}). Shared geometry was therefore not merely unexploited by our decoders; it was undetectable by a measure that uses no decoder.

\begin{figure}[t]
\centering
\includegraphics[width=\textwidth]{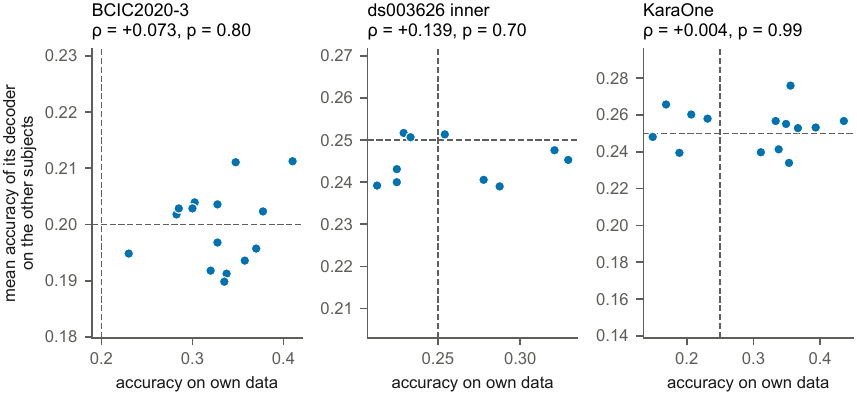}
\caption{Within-subject accuracy of each participant against the mean accuracy of that participant's decoder on every other participant. Each point is one participant; the dashed lines mark chance, and $\rho$ is Spearman's rank correlation.}
\label{fig:source}
\end{figure}

A decoder that barely decodes its own participant cannot show that nothing is shared, only fail to show that something is. We tested this directly by asking whether participants who decode their own data best also produce the decoders that transfer best. They did not (figure~\ref{fig:source}): the Spearman correlation between within-subject accuracy and mean accuracy on other participants was $+$0.073 ($p$ = 0.80) on the competition corpus, $+$0.139 ($p$ = 0.70) on ds003626 and $+$0.004 ($p$ = 0.99) on KaraOne, and correlations with the accuracy that other participants' decoders achieve on each participant were likewise null. The strongest version of the claim rests on the competition corpus, whose within-subject reference lies 0.1273 above chance while transfer is flat. On ds003626 and KaraOne the references lie only 0.0095 and 0.0486 above chance, so those corpora are consistent with the same conclusion but could not establish it on their own.

\subsection{A post-stimulus state generalizes, content does not}\label{sec:state}

On KaraOne, imagery against rest decoded across participants, and unlike the content results it survived its controls. With all 68 stored channels the contrast reached 0.9802 under LOSO with band power, and 0.88 to 0.92 with EEGNet, but most of that accuracy came from non-cortical channels carried by the common average reference \cite{companionA1}. On the 62 scalp channels re-referenced to their own mean it reached 0.6515 (threshold 0.514, 12 of 14 participants, $p$ = 0.00074, 95\% CI [0.612, 0.694]), and no single band reproduced this value.

Rest and imagery come from the same continuous recording, with rest always preceding the stimulus, so the contrast is entangled with elapsed time and slow drift. Varying the separation between rest and imagery windows while keeping the state difference fixed bounds that component (table~\ref{tab:state}). The separated arm reproduced the main result on an independently built dataset (0.6475 against 0.6515). Reducing the separation from 5.0 to 2.0\,s cost 0.0377, 26\% of the margin above chance, and this reduction did not itself reach adjusted significance ($q$ = 0.091). Training on one half of each session and testing on the other cost more and did reach it. The most conservative arm, trained on the first half of each session and tested on the second, retained 0.5776, or 0.078 above chance, with 5 of 14 participants above threshold; the most favorable retained 0.110 above chance with 11 of 14. Roughly a quarter of the effect is therefore attributable to elapsed time and within-session drift.

\begin{table}[t]
\caption{Imagery versus rest on KaraOne under elapsed-time controls (LOSO, re-referenced scalp channels). Windows are relative to stimulus onset. The last column compares each arm with the separated arm; $q$ is the Benjamini--Hochberg adjusted $p$-value.}
\label{tab:state}
\centering
\begin{tabular}{@{}llllccl@{}}
\toprule
Arm & Rest window & Imagery window & Gap & Accuracy & $n_{\geq}$ & Change ($q$) \\
\midrule
Separated & $-$5.0 to $-$3.0\,s & $+$2.0 to $+$4.0\,s & 5.0\,s & 0.6475 & 12/14 & reference \\
Adjacent & $-$2.0 to 0.0\,s & $+$2.0 to $+$4.0\,s & 2.0\,s & 0.6099 & 11/14 & $-$0.0377 (0.091) \\
Late & $-$2.0 to 0.0\,s & $+$5.0 to $+$7.0\,s & 5.0\,s & 0.5970 & 12/14 & $-$0.0505 (0.039) \\
Adjacent, first half to second & & & 2.0\,s & 0.5776 & 5/14 & $-$0.0699 (0.033) \\
Adjacent, second half to first & & & 2.0\,s & 0.6024 & 10/14 & $-$0.0452 (0.039) \\
\bottomrule
\end{tabular}
\end{table}

These controls bound the contribution of elapsed time, but every rest window precedes the stimulus and every imagery window follows it, so they cannot separate imagery from the processing that the stimulus sets off. The pre-imagery interval addresses this (table~\ref{tab:engagement}). On windows cut afresh from the recordings, the adjacent arm reproduced exactly (0.6099). Applied to held-out participants, the rest-versus-imagery decoder labelled 66.3\% of pre-imagery windows as imagery, against 61.4\% of imagery windows and 39.4\% of rest windows. The pre-imagery interval was labelled as imagery more often than rest in all 14 participants ($p$ = 0.00006), whereas imagery was labelled as imagery more often than the pre-imagery interval in only 6 ($p$ = 0.90). Decoded directly, rest and the pre-imagery interval separated at 0.6646, more strongly than rest and imagery, although those two windows are adjacent and differ less in elapsed time.

\begin{table}[!htb]
\caption{Rest, pre-imagery and imagery windows on KaraOne (LOSO, band power, re-referenced scalp channels). The pre-imagery interval lies between rest and imagery and contains perception of the prompt and preparation to articulate it. Above: each pair of windows decoded directly; threshold is the mean of the per-participant thresholds, and each participant is counted in $n_{\geq}$ against their own. Below: fraction of each window type that the rest-versus-imagery decoder, trained on the other participants, labelled as imagery. Each $p$ is a one-sided Wilcoxon test across participants; the shuffle-label controls for the upper rows are in table~\ref{tab:shuffle}.}
\label{tab:engagement}
\centering
\begin{tabular}{@{}lcccl@{}}
\toprule
Contrast & Accuracy & Threshold & $n_{\geq}$ & $p$ \;\; [95\% CI] \\
\midrule
Rest vs imagery & 0.6099 & 0.552 & 11/14 & 0.00006 [0.580, 0.641] \\
Rest vs pre-imagery & 0.6646 & 0.553 & 12/14 & 0.00006 [0.623, 0.706] \\
Pre-imagery vs imagery & 0.6122 & 0.553 & 13/14 & 0.00006 [0.585, 0.647] \\
\midrule
Window & \multicolumn{4}{l}{Labelled as imagery \;\; [95\% CI]} \\
\midrule
Rest & \multicolumn{4}{l}{0.394 [0.290, 0.552]} \\
Pre-imagery & \multicolumn{4}{l}{0.663 [0.545, 0.762]; above rest in 14/14, $p$ = 0.00006} \\
Imagery & \multicolumn{4}{l}{0.614 [0.498, 0.731]; above pre-imagery in 6/14, $p$ = 0.90} \\
\bottomrule
\end{tabular}
\end{table}

The difference from rest is therefore already present before imagery begins, and the contrast does not isolate imagery. The pre-imagery interval and imagery were also separable (0.6122, 13 of 14 participants), but that comparison sets perception of the prompt and preparation to articulate against imagery at a later point in the trial, so it does not isolate imagery either. What generalizes across participants on KaraOne is a change of state that follows the stimulus, at about 0.61 after artifact and elapsed-time controls, and not a signature of imagined speech; speech content does not generalize.

\subsection{Why the within-subject ceiling is low}\label{sec:ceiling}

Under the held-out protocols, the reference decoder reaches 0.3147 on the competition corpus (15 of 15 participants above threshold, $p < 0.0001$), 0.2788 on ds003626 inner speech (2 of 10) and 0.2583 on KaraOne imagined words (1 of 14). We tested five explanations for these values; a change counted as an improvement only if it beat the reference on a paired per-participant test after correction.

The first explanation, a failure upstream of the decoder, was not supported. Applied to the three-way condition contrast of ds003626 (5,640 trials), balanced per participant and session (an unbalanced analysis would reach 0.4035 by predicting the majority class, against a nominal chance of 0.3333), the same pipeline reached 0.6079 with all ten participants above threshold ($p$ = 0.0010) under leave-one-session-out evaluation. Epoching, labels and windows therefore carry decodable structure. Because condition in this corpus is confounded with position within the session \cite{companionA1}, this result shows that the pipeline can read structure the recordings contain, not that it reads speech.

The second, too many features for the available trials, was not supported either. Under the held-out protocols, training folds contained about 265, 150 and 33 rows against 384, 768 and 372 features. Across a family of 20 reductions (a single 8--30\,Hz band with six times fewer features, principal component projections to 16, 32 and 64 components, an anatomical speech-motor region, and cropped training with five times more rows), seven cells reached adjusted significance and all seven were losses. On the competition corpus, projection to 16 components cut the feature count 24-fold and raised the ratio of training rows to features from 0.69 to 16.67, the highest in the grid, and produced the lowest accuracy on that corpus (0 of 15 participants improved, $q$ = 0.013).

The third, that per-channel power misses cross-channel structure, was tested with 23 Riemannian configurations on the same folds. None outperformed band power and four were significantly worse. Contrary to our expectation, the least constrained configuration, tangent-space classification of the full-montage covariance with 2,080 dimensions from 267 training rows, was the best Riemannian configuration on the competition corpus, and every principled reduction was worse; its advantage over band power was not significant (11 of 15 participants, $q$ = 0.40).

\begin{figure}[t]
\centering
\includegraphics[width=\textwidth]{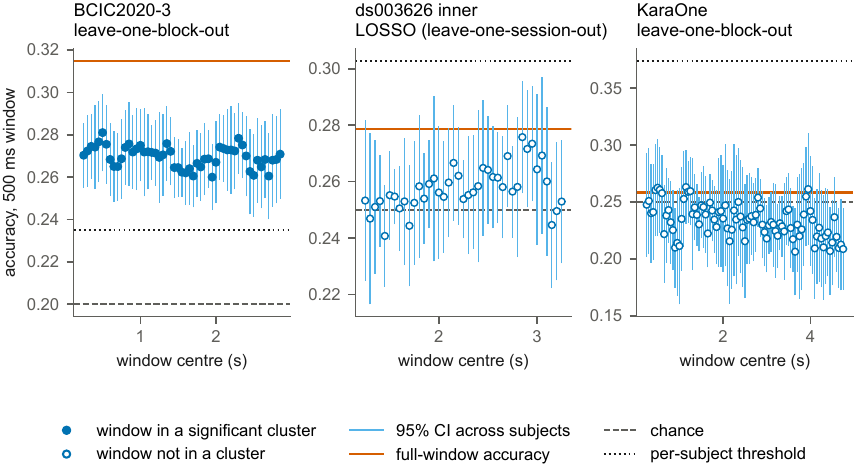}
\caption{Time-resolved decoding with 500\,ms windows in 50\,ms steps. Each point is the mean accuracy over participants in one window, with its 95\% confidence interval; filled points belong to a cluster that survives the cluster-based permutation test over time. The solid line is the accuracy of the full analysis window, the dashed line is chance and the dotted line the per-participant threshold.}
\label{fig:sliding}
\end{figure}

The fourth, that the fixed window dilutes a brief effect, was also not supported (figure~\ref{fig:sliding}). The best 500\,ms window, selected on the data it was scored on and therefore an optimistic estimate, reached 0.2810, 0.2758 and 0.2631 on the three corpora, which is $-$0.0337, $-$0.0030 and $+$0.0048 relative to the full window. On the competition corpus a single cluster spanned the entire epoch (0.00--3.10\,s, $p$ = 0.0001) and every window decoded above chance on its own, so the effect is sustained and integrating over the full trial is appropriate.

The fifth, too little training data, is the only explanation that no change of representation can address, and it is the only variable that orders the corpora: 80 trials per class per participant gives 0.3147 with 15 of 15 participants above threshold, about 56 gives 0.2788 with 2 of 10, and about 12 gives 0.2583 with 1 of 14. It could be tested directly on ds003626, which records the same four words in pronounced and visualized conditions. Adding those trials to the inner-speech training set, restricted to sessions already in the training fold, increased training rows up to 2.5-fold and did not improve accuracy: every arm was nominally lower than the inner-only baseline and none approached significance (supplementary table~S2). Trials of the same words under another condition therefore do not substitute for trials of the target condition, plausibly because the change of condition introduces a shift that costs as much as the extra data provide.

The two Braindecode architectures did not change this picture. Neither decoded ds003626 inner speech under either protocol: ShallowFBCSPNet reached 0.2500 with a session held out and 0.2433 under LOSO, and EEG Conformer 0.2501 and 0.2340, against a chance level of 0.250, with at most one of ten participants above threshold and group $p$ between 0.455 and 0.920. Under LOSO on the competition corpus, ShallowFBCSPNet reached 0.1958 with no participant above threshold. Within subject on that corpus it did decode, at 0.2438 with a block held out (macro-F1 0.2421; 10 of 15 participants above their threshold of 0.235 over 400 test trials; $p$ = 0.0005; 95\% CI 0.2302 to 0.2558), and its shuffle-label control had a null mean of 0.1959 with no draw above 0.2085. On the same test trials band power was higher in 14 of 15 participants, by 0.0708 on average (exact sign-flip test, $p$ = 0.0001). The network reserves a validation set for early stopping and so trains on about 213 rows per fold against about 267 for band power, so the comparison is not matched for training data. On ds003626 ShallowFBCSPNet has 208,324 parameters against about 179 training rows per within-subject fold, which again points to trial count as the binding constraint.

\subsection{Phonological features and held-out prompts}\label{sec:lopo}

Holding out prompts instead of participants asks whether a decoder has learned a phonological feature or has memorized the prompts. A decoder trained on nasality over nine prompts and tested on two it has never seen cannot rely on prompt identity. Table~\ref{tab:lopo} reports all five contrasts.

\begin{table}[t]
\caption{Phonological contrasts on KaraOne with the test prompts held out of training (112 valid hold-out combinations in total). ``All'' uses all 11 prompts; ``phonemes'' uses the seven phonemes only. $q$ is corrected across the family.}
\label{tab:lopo}
\centering
\begin{tabular}{@{}llccccc@{}}
\toprule
Contrast & Prompts & Hold-outs & Accuracy & Threshold & $n_{\geq}$ & $p$ ($q$) \\
\midrule
Nasal & all & 28 & 0.4880 & 0.532 & 1/14 & 0.861 (0.956) \\
 & phonemes & 10 & 0.4980 & 0.556 & 0/14 & 0.725 (0.937) \\
Bilabial & all & 28 & 0.4984 & 0.532 & 2/14 & 0.749 (0.937) \\
 & phonemes & 10 & 0.5082 & 0.556 & 1/14 & 0.155 (0.311) \\
/iy/ & all & 28 & \textbf{0.5883} & 0.532 & 12/14 & \textbf{0.0006 (0.0061)} \\
 & phonemes & 12 & 0.5117 & 0.550 & 4/14 & 0.377 (0.628) \\
/uw/ & all & 18 & 0.4965 & 0.541 & 0/14 & 0.989 (0.989) \\
Consonant/vowel & all & 10 & 0.5121 & 0.556 & 2/14 & 0.121 (0.302) \\
 & phonemes & 10 & 0.5121 & 0.556 & 2/14 & 0.121 (0.302) \\
/iy/, unseen participant & all & 28 & 0.5504 & 0.532 & 5/14 & 0.015 (0.074) \\
\bottomrule
\end{tabular}
\end{table}

Only one cell survived correction, the /iy/ contrast over all 11 prompts (0.5883, 12 of 14 participants above threshold), and its own control accounts for it. Restricted to the seven phonemes, the same contrast gave 0.5117 with 4 of 14 participants and $p$ = 0.38, despite retaining 12 valid hold-outs. The /iy/ contrast is the only one whose positive class consists entirely of phonemes (/diy/, /iy/, /piy/, /tiy/), while 42.9\% of its negative class are phonemes, an imbalance of 57.1 percentage points in prompt type, against 16.7 to 21.4 points for the other contrasts (supplementary table~S3). Since prompt type alone decodes at 0.8371 on KaraOne, a contrast with this imbalance can exceed chance without phonological information, and removing the words removed the effect. The unseen-participant form behaved the same way (0.5504, $q$ = 0.074). No phonological feature in KaraOne generalized to a prompt the decoder had not seen. The /uw/ contrast has a single positive prompt among the phonemes, so its phoneme-only form cannot be constructed and is absent rather than null.

\subsection{Phase, connectivity and coupling}\label{sec:phase}

Every representation above is an amplitude, a band power or a covariance, none of which carries phase. We therefore also decoded content from connectivity, coupling and phase features (table~\ref{tab:phase}). No cell survived correction across the 54 cells, on any corpus, protocol or region set. The largest gain over chance, $+$0.0175, came from the exploratory phase family with an uncorrected $p$ of 0.14 and 3 of 14 participants above threshold. For the two validated families, the companion paper shows that a lagged mixing of 5 to 10\% between the language regions, or a theta-to-beta modulation depth of 5 to 10\%, would have been detected on the same folds and features \cite{companionA1}. Without re-referencing, coupling over the speech-motor region separated KaraOne imagery from rest at 0.7045; every cell reported here uses re-referenced data, for the reason documented in the companion paper.

\begin{table}[t]
\caption{Content decoding from phase, connectivity and coupling features across three corpora, two region sets and both protocols. Benjamini--Hochberg correction is across all 54 cells.}
\label{tab:phase}
\centering
\begin{tabular}{@{}lcccccc@{}}
\toprule
Feature family & Validated & Cells & FDR & Largest gain & Accuracy & $p$ ($q$) \\
\midrule
Connectivity (signed wPLI) & yes & 18 & 0 & $+$0.0165 & 0.2165 & 0.0345 (0.8824) \\
Theta--beta coupling & yes & 18 & 0 & $+$0.0166 & 0.2666 & 0.0176 (0.8824) \\
Phase at a fixed latency & no & 18 & 0 & $+$0.0175 & 0.1603 & 0.1446 (0.9761) \\
\bottomrule
\end{tabular}
\end{table}

\section{Discussion}

Across three corpora, imagined-speech content did not transfer between participants, and the pair-transfer and representational-similarity results indicate why pooled cross-subject decoding fails. If participants shared a class geometry that pooling blurred, some pairs of individuals would transfer, good source participants would transfer better than poor ones, and representational dissimilarity matrices would correlate across participants. None of these held. Source selection therefore has nothing to select from, and alignment or adversarial adaptation lacks the shared class structure it presupposes \cite{zanini2018,he2020,ganin2016}; forcing agreement between representations that share no class structure would be expected to produce negative transfer \cite{zhang2023}. This conclusion is strongest on the competition corpus, where within-subject decoding is robust while transfer is flat, and it applies to the representations computed here. It does not exclude shared structure in a representation we did not compute.

The one contrast that did generalize, imagery against rest on KaraOne, was not specific to imagery. A decoder trained on it labelled the interval before imagery, in which participants perceive the prompt and prepare to articulate it, as imagery as often as imagery itself. The contrast therefore reads a change of state that follows the stimulus. Before a state-versus-rest result is attributed to imagery, it should be compared with a post-stimulus interval that contains no imagery, as was possible here.

The null results are bounded. The companion paper shows that the band-power decoder used as the reference detects an injected mu and beta desynchronization of 0.54\uV\ on ds003626 under a pre-specified criterion, and that the phase features detect lagged mixing and coupling of 5 to 10\% \cite{companionA1}. The same study identifies what the bounds do not cover: amplitude- and covariance-based decoders did not detect class-dependent phase locking at any injected size, and an effect whose phase is tied to an internal event with variable timing would not be captured by phase measured at a fixed latency. A reader who suspects that imagined-speech content is present in these recordings would need to locate it in an effect of those types, or below those amplitudes.

The within-subject ceiling is low for a reason that no representation addressed. Reducing dimensionality lowered accuracy, including on the corpus with the most statistical power, covariance features were indistinguishable from band power, shorter windows did not help, and two published deep architectures did not exceed the linear baseline. Two of these outcomes were the opposite of our expectations: the most constrained feature set performed worst, and the least constrained Riemannian configuration performed best. The one variable that ordered the three corpora was the number of trials per class, from 12 to 80 per participant, and adding trials from other conditions of the same words did not substitute for it. The most direct route to a better within-subject ceiling is therefore more trials of the target condition per participant, or methods that extract more from few trials than any tested here.

Our results are consistent with the independent re-analyses. On the competition corpus, our within-subject figures of 0.2980 to 0.3273 with sessions pooled (8--30\,Hz band power and the six-band reference, respectively) and 0.3147 with a block held out are comparable to or above the 25.7 to 30.2\% reported for five models under a within-subject protocol, and both studies find chance under LOSO \cite{khanday2026}. They are also consistent with the conclusion that within-category imagined-speech decoding is mostly at or near chance \cite{vandyck2026}. The much higher accuracies reported in earlier work on these corpora \cite{zhao2015,saha2019,jeong2022} were obtained with different protocols, feature engineering and task framings, and we cannot fully account for the difference. Our own strongest binary result on KaraOne, the /iy/ contrast over all 11 prompts (0.5883 with the test prompts held out, table~\ref{tab:lopo}), was explained by prompt type, and binary KaraOne results reported without a phoneme-only control may reflect the same confound.

The study has several limitations. Within-subject figures with sessions pooled are optimistic: holding out a session or run changed the ceiling by $+$0.0162 on average across six corpus and feature-set combinations, with the largest reduction, 0.0402, on KaraOne (supplementary table~S5); for ds003626 inner speech the change was small or reversed, so that null does not depend on session leakage. The analysis windows were fixed per corpus, although time-resolved decoding showed no better window. The deep architectures were trained from scratch on each corpus at their published settings without tuning, so they test architecture and not scale, and we did not evaluate foundation models pretrained on large EEG collections. Finally, no window in these paradigms isolates imagery: the interval before imagery on KaraOne combines perception of the prompt with preparation to articulate it, so the comparison of that interval with imagery is confounded with both and with time in the trial.

\section{Conclusions}

Under a leakage-controlled protocol on three public corpora, imagined-speech content recorded with scalp EEG did not transfer between individual participants, no shared representational geometry was detectable without a classifier, and subject-level alignment did not change this. A contrast between imagery and rest did generalize, but the interval before imagery carried the same distinction, so it reflects a state that follows the stimulus rather than imagery. Within subject, accuracy was above chance for every participant on only one corpus, where it did not pass an artifact audit; of five explanations for the low ceiling, four were excluded by experiment, and the fifth, the number of trials per class, was the only variable that ordered the corpora. These results locate the main obstacle to calibration-free imagined-speech decoding from the scalp in the signal and the amount of data per condition rather than in the choice of decoder.

\ack{During the preparation of this work the authors used Claude (Anthropic) to assist with revising analysis and figure code, and with editing the manuscript text. The authors reviewed, validated and critically assessed all AI-assisted output, checked every reference against its source, and take full responsibility for the content of this article.
}

\funding{This research received no specific grant from any funding agency in the public, commercial or not-for-profit sectors.}

\roles{Blessing Isoyiza Adeika: conceptualization, methodology, software, formal analysis, investigation, data curation, visualization, writing -- original draft. Onyema Osuagwu: conceptualization, methodology, supervision, writing -- review and editing.}

\data{All three corpora are publicly available. ds003626 is available from OpenNeuro \cite{nieto2022data,markiewicz2021}, KaraOne from the University of Toronto \cite{zhao2015}, and the 2020 BCI Competition data from the competition organizers \cite{jeong2022}. Analysis code, per-run records, split hashes and the scripts that generate every table and figure are available at \url{https://github.com/blessadeik/scalp-eeg-speech-decoding}.
}

\suppdata{Supplementary material (PDF): run identifiers for every reported cell (tables~S1a and S1b); multi-condition training on ds003626 (table~S2); prompt-type composition of the phonological contrasts (table~S3); balanced accuracy and exact permutation tests for the content cells (table~S4); the effect of session leakage on the within-subject ceiling (table~S5); the syllable-count control on the competition corpus (table~S6); and run identifiers for the KaraOne pre-imagery analysis (table~S7).}

\section*{Conflict of interest}
The authors declare no conflicts of interest.

\section*{Ethical statement}
This study analyzes only publicly released, de-identified datasets whose collection was approved by the ethics boards named in the original publications \cite{nieto2022,zhao2015,jeong2022}. No new data were collected, and no further approval was required for this secondary analysis.

\bibliographystyle{jne_iop_num}
\bibliography{references}

\clearpage
\section*{Supplementary material}
\setcounter{table}{0}
\renewcommand{\thetable}{S\arabic{table}}

\section*{S1. Run identifiers}
\renewcommand{\thetable}{S1\alph{table}}
Each identifier names an append-only record released with the analysis code, containing the run's configuration, per-fold split hashes and per-participant results. Identifiers carry the dataset's task label (\texttt{inner\_speech} for every ds003626 condition), not the condition; the condition of each run is recorded inside its record.

\begin{table}[H]
\caption{Run identifiers for the content cells of table~3 of the main text. Accuracy is repeated so that rows can be matched.}
\label{tab:s1a}
\centering
\begin{tabular}{@{}lllcp{50mm}@{}}
\toprule
Task & Protocol & Decoder & Accuracy & Run identifier \\
\midrule
Competition, 5 phrases & 5-fold & band power & 0.3273 & \texttt{\small 20260820\_\allowbreak 140120\_\allowbreak bandpower\_\allowbreak lr\_\allowbreak bcic2020\_\allowbreak within-subject-5fold} \\
Competition, 5 phrases & held out & band power & 0.3147 & \texttt{\small 20260821\_\allowbreak 135506\_\allowbreak bandpower\_\allowbreak lr\_\allowbreak bcic2020\_\allowbreak leave-one-block-out} \\
Competition, 5 phrases & LOSO & band power & 0.2063 & \texttt{\small 20260820\_\allowbreak 140122\_\allowbreak bandpower\_\allowbreak lr\_\allowbreak bcic2020\_\allowbreak LOSO} \\
ds003626, 4 words, inner & 5-fold & band power & 0.2595 & \texttt{\small 20260802\_\allowbreak 002846\_\allowbreak bandpower\_\allowbreak lr\_\allowbreak inner\_\allowbreak speech\_\allowbreak within-subject-5fold} \\
ds003626, 4 words, inner & held out & band power & 0.2786 & \texttt{\small 20260821\_\allowbreak 151002\_\allowbreak bandpower\_\allowbreak lr\_\allowbreak inner\_\allowbreak speech\_\allowbreak LOSSO-(leave-one-session-out)} \\
ds003626, 4 words, inner & LOSO & band power & 0.2590 & \texttt{\small 20260802\_\allowbreak 002905\_\allowbreak bandpower\_\allowbreak lr\_\allowbreak inner\_\allowbreak speech\_\allowbreak LOSO} \\
ds003626, 4 words, pronounced & 5-fold & EEGNet & 0.4231 & \texttt{\small 20261001\_\allowbreak 103330\_\allowbreak eegnet\_\allowbreak inner\_\allowbreak speech\_\allowbreak within-subject-5fold\_\_2} \\
ds003626, 4 words, pronounced & LOSO & EEGNet & 0.2511 & \texttt{\small 20260818\_\allowbreak 162813\_\allowbreak eegnet\_\allowbreak inner\_\allowbreak speech\_\allowbreak LOSO} \\
KaraOne, 4 words, imagined & 5-fold & band power & 0.2985 & \texttt{\small 20260821\_\allowbreak 130640\_\allowbreak bandpower\_\allowbreak lr\_\allowbreak karaone\_\allowbreak within-subject-5fold} \\
KaraOne, 4 words, imagined & held out & band power & 0.2583 & \texttt{\small 20260821\_\allowbreak 130641\_\allowbreak bandpower\_\allowbreak lr\_\allowbreak karaone\_\allowbreak leave-one-block-out} \\
KaraOne, 4 words, imagined & LOSO & band power & 0.2643 & \texttt{\small 20260821\_\allowbreak 130641\_\allowbreak bandpower\_\allowbreak lr\_\allowbreak karaone\_\allowbreak LOSO} \\
KaraOne, 7 phonemes, imagined & 5-fold & phase & 0.1603 & \texttt{\small 20260903\_\allowbreak 170441\_\allowbreak phase\_\allowbreak lr\_\allowbreak karaone\_\allowbreak within-subject-5fold} \\
KaraOne, 7 phonemes, imagined & LOSO & phase & 0.1570 & \texttt{\small 20260903\_\allowbreak 170517\_\allowbreak phase\_\allowbreak lr\_\allowbreak karaone\_\allowbreak LOSO} \\
KaraOne, 7 phonemes, pre-imagery & 5-fold & EEGNet & 0.1666 & \texttt{\small 20261005\_\allowbreak 151504\_\allowbreak eegnet\_\allowbreak karaone\_\allowbreak within-subject-5fold} \\
KaraOne, 7 phonemes, pre-imagery & LOSO & EEGNet & 0.1661 & \texttt{\small 20261005\_\allowbreak 153029\_\allowbreak eegnet\_\allowbreak karaone\_\allowbreak LOSO} \\
\bottomrule
\end{tabular}
\end{table}

\begin{table}[H]
\caption{Run identifiers for the alignment cells of table~5 of the main text.}
\label{tab:s1b}
\centering
\begin{tabular}{@{}llcp{66mm}@{}}
\toprule
Corpus & Alignment & Accuracy & Run identifier \\
\midrule
Competition & none & 0.2080 & \texttt{\small 20260820\_\allowbreak 161905\_\allowbreak bandpower\_\allowbreak lr\_\allowbreak bcic2020\_\allowbreak LOSO} \\
Competition & Euclidean, inductive 10 & 0.2063 & \texttt{\small 20260820\_\allowbreak 162013\_\allowbreak bandpower\_\allowbreak lr\_\allowbreak bcic2020\_\allowbreak LOSO} \\
Competition & Euclidean, inductive 20 & 0.2122 & \texttt{\small 20260820\_\allowbreak 162038\_\allowbreak bandpower\_\allowbreak lr\_\allowbreak bcic2020\_\allowbreak LOSO} \\
Competition & Euclidean, inductive 40 & 0.2055 & \texttt{\small 20260820\_\allowbreak 162103\_\allowbreak bandpower\_\allowbreak lr\_\allowbreak bcic2020\_\allowbreak LOSO} \\
Competition & Euclidean, transductive & 0.2013 & \texttt{\small 20260820\_\allowbreak 161933\_\allowbreak bandpower\_\allowbreak lr\_\allowbreak bcic2020\_\allowbreak LOSO} \\
Competition & Riemannian & 0.2052 & \texttt{\small 20260820\_\allowbreak 162202\_\allowbreak bandpower\_\allowbreak lr\_\allowbreak bcic2020\_\allowbreak LOSO} \\
ds003626 inner & none & 0.2453 & \texttt{\small 20260820\_\allowbreak 162220\_\allowbreak bandpower\_\allowbreak lr\_\allowbreak inner\_\allowbreak speech\_\allowbreak LOSO} \\
ds003626 inner & Euclidean, inductive 10 & 0.2402 & \texttt{\small 20260820\_\allowbreak 162250\_\allowbreak bandpower\_\allowbreak lr\_\allowbreak inner\_\allowbreak speech\_\allowbreak LOSO} \\
ds003626 inner & Euclidean, inductive 20 & 0.2440 & \texttt{\small 20260820\_\allowbreak 162305\_\allowbreak bandpower\_\allowbreak lr\_\allowbreak inner\_\allowbreak speech\_\allowbreak LOSO} \\
ds003626 inner & Euclidean, inductive 40 & 0.2490 & \texttt{\small 20260820\_\allowbreak 162319\_\allowbreak bandpower\_\allowbreak lr\_\allowbreak inner\_\allowbreak speech\_\allowbreak LOSO} \\
ds003626 inner & Euclidean, transductive & 0.2539 & \texttt{\small 20260820\_\allowbreak 162237\_\allowbreak bandpower\_\allowbreak lr\_\allowbreak inner\_\allowbreak speech\_\allowbreak LOSO} \\
ds003626 inner & Riemannian & 0.2462 & \texttt{\small 20260820\_\allowbreak 162641\_\allowbreak bandpower\_\allowbreak lr\_\allowbreak inner\_\allowbreak speech\_\allowbreak LOSO} \\
KaraOne & none & 0.2390 & \texttt{\small 20260820\_\allowbreak 162647\_\allowbreak bandpower\_\allowbreak lr\_\allowbreak karaone\_\allowbreak LOSO} \\
KaraOne & Euclidean, inductive 10 & 0.2771 & \texttt{\small 20260820\_\allowbreak 162656\_\allowbreak bandpower\_\allowbreak lr\_\allowbreak karaone\_\allowbreak LOSO} \\
KaraOne & Euclidean, inductive 20 & 0.2631 & \texttt{\small 20260820\_\allowbreak 162700\_\allowbreak bandpower\_\allowbreak lr\_\allowbreak karaone\_\allowbreak LOSO} \\
KaraOne & Euclidean, inductive 40 & 0.2753 & \texttt{\small 20260820\_\allowbreak 162704\_\allowbreak bandpower\_\allowbreak lr\_\allowbreak karaone\_\allowbreak LOSO} \\
KaraOne & Euclidean, transductive & 0.2780 & \texttt{\small 20260820\_\allowbreak 162652\_\allowbreak bandpower\_\allowbreak lr\_\allowbreak karaone\_\allowbreak LOSO} \\
KaraOne & Riemannian & 0.2771 & \texttt{\small 20260820\_\allowbreak 162711\_\allowbreak bandpower\_\allowbreak lr\_\allowbreak karaone\_\allowbreak LOSO} \\
\bottomrule
\end{tabular}
\end{table}

\section*{S2. Multi-condition training on ds003626}
\setcounter{table}{1}
\renewcommand{\thetable}{S\arabic{table}}
The test set contains inner-speech trials only, and the split is drawn on inner trials alone, so it is identical to the baseline's. Additional trials were taken only from sessions already present in the training fold; adding them from the held-out session would have supplied that session's electrode placement and drift. Arms were run separately because the pronounced condition is dominated by muscle activity (companion paper), so any gain would have needed to be attributed to a source.

\begin{table}[H]
\caption{Inner-speech accuracy on ds003626 (held-out session) when pronounced and visualized trials of the same words are added to training. $q$ is from a paired per-participant test with Benjamini--Hochberg correction.}
\label{tab:s2}
\centering
\begin{tabular}{@{}lcccccc@{}}
\toprule
Training set & Rows per fold & Multiplier & Accuracy & Change & $n_{\geq}$ & $q$ \\
\midrule
inner only & 149 & 1.00 & 0.2788 & reference & 2/10 & n/a \\
+ pronounced & 224 & 1.50 & 0.2765 & $-$0.0023 & 2/10 & 1.000 \\
+ visualized & 301 & 2.02 & 0.2672 & $-$0.0116 & 1/10 & 0.463 \\
+ both & 376 & 2.52 & 0.2705 & $-$0.0083 & 0/10 & 0.648 \\
\bottomrule
\end{tabular}
\end{table}

\section*{S3. Prompt-type composition of the phonological contrasts}

\begin{table}[H]
\caption{Proportion of phoneme prompts (as opposed to word prompts) in each class of the KaraOne contrasts, and the difference in percentage points.}
\label{tab:s3}
\centering
\begin{tabular}{@{}lccc@{}}
\toprule
Contrast & Positive class & Negative class & Difference \\
\midrule
/iy/ & 100\% & 42.9\% & 57.1 \\
nasal & 50.0\% & 71.4\% & 21.4 \\
bilabial & 50.0\% & 71.4\% & 21.4 \\
/uw/ & 50.0\% & 66.7\% & 16.7 \\
\bottomrule
\end{tabular}
\end{table}

\section*{S4. Balanced accuracy and exact permutation tests}
Balanced accuracy is macro-averaged recall and macro-F1 the unweighted mean of per-class F1 scores, both computed from each participant's summed confusion matrix. The permutation $p$ is an exact one-sided sign-flip test on the per-participant departures from chance, enumerating all $2^n$ sign assignments; it complements the Wilcoxon test, which ranks those departures and discards their magnitude. No configuration in this table uses the held-out participant's data in any form. Accuracy here is averaged over folds, which differs from the mean over participants in table~3 of the main text only for ds003626 inner speech with a session held out (0.2786 against 0.2788).

\begin{table}[H]
\caption{Content cells of table~3 of the main text under class-balanced metrics and an exact permutation test.}
\label{tab:s4}
\centering
\begin{tabular}{@{}lllcccc@{}}
\toprule
Task & Protocol & Decoder & Accuracy & Balanced & Macro-F1 & Perm.\ $p$ \\
\midrule
Competition, 5 phrases & 5-fold & band power & 0.3273 & 0.3273 & 0.3270 & 0.0001 \\
Competition, 5 phrases & held out & band power & 0.3147 & 0.3147 & 0.3142 & 0.0001 \\
Competition, 5 phrases & LOSO & band power & 0.2063 & 0.2063 & 0.1796 & 0.2234 \\
ds003626, 4 words, inner & 5-fold & band power & 0.2595 & 0.2595 & 0.2595 & 0.2439 \\
ds003626, 4 words, inner & held out & band power & 0.2786 & 0.2788 & 0.2729 & 0.0127 \\
ds003626, 4 words, inner & LOSO & band power & 0.2590 & 0.2590 & 0.2116 & 0.0898 \\
ds003626, 4 words, pronounced & 5-fold & EEGNet & 0.4231 & 0.4231 & 0.4178 & 0.0020 \\
ds003626, 4 words, pronounced & LOSO & EEGNet & 0.2511 & 0.2511 & 0.2466 & 0.4790 \\
KaraOne, 4 words, imagined & 5-fold & band power & 0.2985 & 0.2985 & 0.2992 & 0.0383 \\
KaraOne, 4 words, imagined & held out & band power & 0.2583 & 0.2583 & 0.2490 & 0.3291 \\
KaraOne, 4 words, imagined & LOSO & band power & 0.2643 & 0.2643 & 0.2088 & 0.1719 \\
KaraOne, 7 phonemes, imagined & 5-fold & phase & 0.1603 & 0.1603 & 0.1587 & 0.0926 \\
KaraOne, 7 phonemes, imagined & LOSO & phase & 0.1570 & 0.1570 & 0.1524 & 0.0711 \\
KaraOne, 7 phonemes, pre-imagery & 5-fold & EEGNet & 0.1666 & 0.1659 & 0.1594 & 0.0527 \\
KaraOne, 7 phonemes, pre-imagery & LOSO & EEGNet & 0.1661 & 0.1655 & 0.1518 & 0.0500 \\
\bottomrule
\end{tabular}
\end{table}

\section*{S5. Session leakage and the within-subject ceiling}
The held-out protocol tests on an unseen session for ds003626, an unseen run for the competition corpus (whose three runs are the competition's 300/50/50 partition rather than separate recording days) and an unseen contiguous block for KaraOne, which has one session. The competition corpus's leave-one-block-out figure in the main text (0.3147) uses balanced blocks and differs from the leave-one-run-out figure here. Averaged over the six rows, pooling sessions raised within-subject accuracy by 0.0162.

\begin{table}[H]
\caption{Within-subject accuracy with sessions pooled (5-fold) and with a session, run or block held out, for the two band-power feature sets.}
\label{tab:s5}
\centering
\begin{tabular}{@{}llccc@{}}
\toprule
Corpus & Features & 5-fold & Held out & Pooled minus held out \\
\midrule
Competition & six-band bank & 0.3273 & 0.2952 & $+$0.0322 \\
Competition & 8--30\,Hz & 0.2980 & 0.2683 & $+$0.0297 \\
ds003626 inner & six-band bank & 0.2595 & 0.2788 & $-$0.0193 \\
ds003626 inner & 8--30\,Hz & 0.2537 & 0.2513 & $+$0.0024 \\
KaraOne & six-band bank & 0.2985 & 0.2583 & $+$0.0402 \\
KaraOne & 8--30\,Hz & 0.2622 & 0.2500 & $+$0.0122 \\
\bottomrule
\end{tabular}
\end{table}

\section*{S6. Syllable-count control on the competition corpus}
``Stop'' and ``yes'' have one syllable and ``hello'', ``help me'' and ``thank you'' two, so four of the ten phrase pairs are matched for length and six are not. Each participant's accuracy was averaged over the matched and over the mismatched pairs, using the pairwise cells of the main text at 256\,Hz (80 logged runs, one per pair, band and protocol, gathered in the record \texttt{\small per\_word\_cells.json}). No model was retrained. The test of a length effect is a one-sided paired Wilcoxon test of mismatched against matched pairs across the 15 participants; $q$ is Benjamini--Hochberg adjusted across the four bands within each protocol.

\begin{table}[H]
\caption{Pairwise separability on the competition corpus for phrase pairs matched and mismatched in syllable count. Chance is 0.500. ``Within'' is 5-fold cross-validation within subject. The last two columns test the matched pairs alone against chance.}
\label{tab:s6}
\centering
\setlength{\tabcolsep}{4pt}
\begin{tabular}{@{}llcccccl@{}}
\toprule
Protocol & Band & Matched & Mismatched & Difference & $p$ ($q$) & Matched $p$ & Matched 95\% CI \\
\midrule
within & 8--13\,Hz & 0.5483 & 0.5634 & $+$0.0151 & 0.019 (0.025) & 0.0007 & [0.527, 0.561] \\
within & 8--30\,Hz & 0.5731 & 0.5932 & $+$0.0201 & 0.008 (0.025) & 0.0004 & [0.550, 0.589] \\
within & 62--100\,Hz & 0.6673 & 0.6827 & $+$0.0154 & 0.084 (0.084) & 0.0003 & [0.648, 0.686] \\
within & 8--30 + 62--100\,Hz & 0.6566 & 0.6785 & $+$0.0220 & 0.013 (0.025) & $<$0.0001 & [0.636, 0.673] \\
LOSO & 8--13\,Hz & 0.5051 & 0.5156 & $+$0.0105 & 0.115 & 0.165 & [0.494, 0.517] \\
LOSO & 8--30\,Hz & 0.5063 & 0.5144 & $+$0.0082 & 0.211 & 0.266 & [0.494, 0.519] \\
LOSO & 62--100\,Hz & 0.5054 & 0.5033 & $-$0.0021 & 0.598 & 0.160 & [0.488, 0.518] \\
LOSO & 8--30 + 62--100\,Hz & 0.5042 & 0.5142 & $+$0.0101 & 0.180 & 0.257 & [0.488, 0.517] \\
\bottomrule
\end{tabular}
\end{table}

\section*{S7. Run identifiers for the KaraOne pre-imagery analysis}

\begin{table}[H]
\caption{Run identifiers for the three contrasts of table~8 of the main text. Each run carries its own shuffle-label control of 200 draws. The transfer rates in the lower half of that table were computed on the same participant folds as the rest-versus-imagery run.}
\label{tab:s7}
\centering
\begin{tabular}{@{}lcp{80mm}@{}}
\toprule
Contrast & Accuracy & Run identifier \\
\midrule
Rest vs imagery & 0.6099 & \texttt{\small 20260930\_\allowbreak 175026\_\allowbreak bandpower\_\allowbreak lr\_\allowbreak karaone\_\allowbreak LOSO} \\
Rest vs pre-imagery & 0.6646 & \texttt{\small 20260930\_\allowbreak 182642\_\allowbreak bandpower\_\allowbreak lr\_\allowbreak karaone\_\allowbreak LOSO} \\
Pre-imagery vs imagery & 0.6122 & \texttt{\small 20260930\_\allowbreak 190302\_\allowbreak bandpower\_\allowbreak lr\_\allowbreak karaone\_\allowbreak LOSO} \\
\bottomrule
\end{tabular}
\end{table}

\end{document}